\documentclass[twocolumn]{aa}
\usepackage{psfig}
\usepackage{natbib}
\bibpunct{(}{)}{;}{a}{}{,}
\usepackage{ltxtable}
\usepackage[latin1]{inputenc}
\begin{document}
   \title{Temporal evolution of magnetic molecular shocks}
  \subtitle{II. Analytics of the steady state and\\
semi-analytical construction of intermediate ages.}

   \author{P. Lesaffre \inst{1,2,3}         
        \and
        J.-P. Chièze \inst{3}	
	\and
	S. Cabrit \inst{4}
	\and
	G. Pineau des Forêts \inst{5,6}
	  }

   \offprints{lesaffre@ast.cam.ac.uk}

   \institute{Institute of Astronomy, Madingley Road, Cambridge 
CB3~0HA, UK 
\and University of Oxford, Department of Astrophysics,
 Oxford OX1~3RH, UK
\and CEA/DAPNIA/SAp Orme des Merisiers, F-91191
   Gif-sur-Yvette Cedex 
\and LERMA, UMR 8112 du CNRS, Observatoire de
   Paris, 61 Av. de l'Observatoire, F-75014 Paris 
\and IAS, UMR-8617 du CNRS, Université Paris-Sud, bât. 121, F-91405 Orsay 
\and LUTH,
   UMR-8102 du CNRS, Observatoire de Paris, F-92190 Meudon Cedex }
   \date{Received December 15, 2003; Accepted June 10, 2004}

   \abstract{In the first paper of this series (Paper~I)
 we computed time dependent
   simulations of multifluid shocks with chemistry and a transverse
   magnetic field frozen in the ions, using an adaptive moving grid.

In this paper, we present new analytical results on steady-state
molecular shocks. Relationships between density and
pressure in the neutral fluid are derived for the cold magnetic
precursor, hot magnetic precursor, adiabatic shock front, and the
following cooling layer. The compression ratio and temperature behind
a fully dissociative adiabatic shock is also derived.

To prove that these results may even hold for intermediate ages, we
design a test to locally characterise the validity of the steady state
equations in a time-dependent shock simulation. Applying this tool to
the results of Paper~I, we show that most of these shocks (all the
stable ones) are indeed in a quasi-steady state at all times, i.e.~: a
given snapshot is composed of one or more truncated steady
shock. Finally, we use this property to produce a construction method
of any intermediate time of low velocity shocks ($u < 20$~kms$^{-1}$) with
only a steady-state code. In particular, this method
allows one to predict the occurrence of steady CJ-type shocks more
accurately than previously proposed criteria.

   }

   \maketitle

\section{Introduction}

  In a previous paper \citep[][ hereafter
   Paper~I]{Les04}(\citet{Les04}, we presented a numerical method to
  compute the time-dependent evolution of molecular shocks with a
  realistic cooling and chemistry in the presence of a transverse magnetic
  field. The use of a moving grid algorithm allowed us to reduce the
  number of zones and the computational time. However, the computation
  of the evolution of a stable shock from formation to steady-state still
  involves one or two days of CPU time on a 500 MHz
  workstation. Oscillating shocks require one week or two. This
  prevents the use of this code to fit shock parameters to
  observations. 

On the other hand, steady-state codes give a fast answer (in one or
two minutes) and include much richer physics. They have therefore been
extensively used to interpret observed spectra. Nevertheless,
steady-state codes have their limits. Observed magnetic molecular
shocks may not be fully in steady-state yet, in which case they show a
combination of C-type and J-type features. They might also be mildly
or strongly unstable \citep[][ and Paper~I of this
series]{Lim02,SR03}, in which case a steady state model will have
limited success.

However, different attempts have been made to
circumvent these problems.  In the field of shocks in supernovae
remnants, \citet{R88} among others were successful in interpreting
spectra with truncated steady J-shocks. Those models allowed them to
account for incomplete recombination zones in a filament of the Cygnus
loop. \citet{CPF98} discovered that the nascent magnetic precursor in
a C-type shock was identical to a truncation of the steady state shock
(for sufficiently late ages). \citet{FP99} used this result to
reproduce H$_2$ excitation diagrams in Cepheus A West.

The aim of this paper is to rigorously test the ideas of
\citet{R88} and \citet{FP99}, by clarifying the
relationship between time-dependent models and steady-state models.
Indeed, we will show that time-dependence is within
reach of steady state codes, as long as the shock is not subject to
strong instabilities.

In Sect. 2 we study the stationary equations of magnetic shocks,
and derive new analytical relations. Section 2 is independent of the
other sections. Then, in Sect. 3, we assess the
validity of the stationary approach by using the results of our fully
hydrodynamical code (Paper~I). In Sect. 4, we explain how to build
time-dependent models of low velocity shocks with the only help of a
steady-state code. We discuss our results in Sect. 5 and sum up our
conclusions in Sect. 6.

  
\section{Analytics of steady shocks}

\label{derivationsteady}

When the flow is in a steady state, time derivatives
 can be skipped in the steady frame. We derive here a few
analytical relations valid along such steady flows.

\subsection{Dynamical equations in a conservative form}

  We recall here the time-dependent monodimensional equations of
  multifluid hydrodynamics with a frozen transverse magnetic
  field. We put them in their conservative form~:

\begin{equation}
\label{hydro1}
\frac{\partial }{\partial t}(n_j)+\frac{\partial}{\partial x}(n_j u_{\rm n}+J_j)=
R_j \mbox{ for $j$ neutral specie }
\end{equation} 

\begin{equation}
\frac{\partial }{\partial t}(n_j)+\frac{\partial}{\partial x}(n_j u_{\rm c}+J_j)=
R_j \mbox{ for $j$ ionic specie }
\end{equation} 

\begin{equation}
\frac{\partial}{\partial t}(\rho_{\rm n} u_{\rm n})+\frac{\partial}{\partial x}(\rho_{\rm n} u_{\rm n}^2+p_{\rm n}+\pi_{\rm n})=
F_{\rm c\rightarrow n}
\end{equation}

\begin{equation}
\frac{\partial}{\partial t}(\rho_{\rm c} u_{\rm c})+
\frac{\partial}{\partial x}(\rho_{\rm c} u_{\rm c}^2+p_{\rm c}+\pi_{\rm i}+\frac{B^2}{8 \pi})=
F_{\rm n\rightarrow c}
\end{equation} 

\begin{displaymath}
\frac{\partial}{\partial t}(\frac1{\gamma-1}p_{\rm n}+\frac12\rho_{\rm n}u_{\rm n}^2)+
\frac{\partial}{\partial x}[u_{\rm n} (\frac{\gamma}{\gamma-1} p_{\rm n}+\frac12\rho_{\rm n}u_{\rm n}^2+\pi_{\rm n})]
\end{displaymath}
\begin{equation}
=\Lambda_{\rm n}+Q_{\rm i\rightarrow n}+Q_{\rm e\rightarrow n}
+u_{\rm n}F_{\rm c\rightarrow n}-\frac12u_{\rm n}^2M_{\rm n}
\end{equation} 

\begin{displaymath}
\frac{\partial}{\partial t}(\frac1{\gamma-1}p_{\rm i}+\frac12\rho_{\rm i}u_{\rm c}^2+\frac{B^2}{8\pi})
\end{displaymath}
\begin{displaymath}
+
\frac{\partial}{\partial x}[u_{\rm c} (\frac{\gamma}{\gamma-1} p_{\rm i}+\frac12\rho_{\rm i}u_{\rm c}^2+\pi_{\rm i}+\frac{B^2}{
4\pi})]
\end{displaymath}
\begin{equation}
=\Lambda_{\rm i}+Q_{\rm n\rightarrow i}+Q_{\rm e\rightarrow i}
+u_{\rm c}F_{\rm n\rightarrow c}-\frac12u_{\rm c}^2M_{\rm i}
\end{equation} 

\begin{displaymath}
\frac{\partial}{\partial t}(\frac1{\gamma-1}p_{\rm e}+\frac12\rho_{\rm e}u_{\rm c}^2)+
\frac{\partial}{\partial x}[u_{\rm c} (\frac{\gamma}{\gamma-1} p_{\rm e}+\frac12\rho_{\rm e}u_{\rm c}^2)]
\end{displaymath}
\begin{equation}
=\Lambda_{\rm e}+Q_{\rm n\rightarrow e}+Q_{\rm i\rightarrow e}
-\frac12u_{\rm c}^2M_{\rm e}
\end{equation} 

\begin{equation}
\frac{\partial}{\partial t}(B)+\frac{\partial}{\partial x}(u_{\rm c} B)=0
\label{hydro2}
\end{equation}
   n, i, e, and c
  indices stand for neutrals, ions, electrons, and charges. $n_j$,
  $\rho$, $u$, $B$, $p$, and $\pi$ are respectively the number
  densities, mass densities, velocities, magnetic field,
  thermal and viscous pressures. $M$, $F$, $Q$ are the mass, momentum, and
  heat transfer rates. $\Lambda$ denotes radiative losses. $R_j$
  stands for chemical rates, and $J_j$ for diffusive fluxes.

\subsection{Steady state equations}
  Let us assume we are in a frame where the flow is in a steady state.
We may then drop the $\frac{\partial}{\partial t}$ terms in equations
\ref{hydro1}-\ref{hydro2}.  If we now integrate equations \ref{hydro1}-\ref{hydro2} along
the $x$ coordinate, we link the state of the gas at one point $x$ in the
shock to the state of the gas far upstream, i.e. to the entrance
parameters of the shock (denoted with a 0 superscript in the following). We give
here the result of such an integration in terms of conserved
fluxes (dotted letters) through the steady region :

 Mass flux~:
\begin{equation}
\dot{M_{\rm n}}=\rho^0_{\rm n} u^0_{\rm n}=\rho_{\rm n} u_{\rm n}-\int_0^{x}M_{\rm n}{\rm d}x'
\end{equation}

\begin{equation}
\dot{M_{\rm c}}=\rho^0_{\rm c} u^0_{\rm c}=\rho_{\rm c} u_{\rm c}+\int_0^{x}M_{\rm n}{\rm d}x'
\end{equation}

 Momentum flux~:
\begin{displaymath}
\dot{P_{\rm n}}=\rho^0_{\rm n} (u^0_{\rm n})^2+p^0_{\rm n}
\end{displaymath}
\begin{equation}
=\rho_{\rm n} u_{\rm n}^2+p_{\rm n}+\pi_{\rm n}-\int_0^{x}F_{\rm c\rightarrow n}{\rm d}x'
\end{equation}

\begin{displaymath}
\dot{P_{\rm c}}=\rho^0_{\rm c} (u^0_{\rm c})^2+p^0_{\rm c}+\frac{(B^0)^2}{8 \pi}
\end{displaymath}
\begin{equation}
=\rho_{\rm c} u_{\rm c}^2+p_{\rm c}+\pi_{\rm i}+\frac{B^2}{8 \pi}-\int_0^{x}F_{\rm n\rightarrow c}{\rm d}x'
\end{equation}

 Energy flux~:
\begin{displaymath}
\dot{E_{\rm n}}=u^0_{\rm n} (\frac{\gamma}{\gamma-1} p^0_{\rm n}+\frac12\rho^0_{\rm n} (u^0_{\rm n})^2)
\end{displaymath}
\begin{equation}
=u_{\rm n} (\frac{\gamma}{\gamma-1} p_{\rm n}+\frac12\rho_{\rm n} u_{\rm n}^2+\pi_{\rm n})
-\int_0^{x}S_{\rm n}{\rm d}x'
\end{equation}

\begin{displaymath}
\dot{E_{\rm i}}=u^0_{\rm c} (\frac{\gamma}{\gamma-1} p^0_{\rm i}+\frac12\rho^0_{\rm i} (u^0_{\rm c})^2+\frac{(B^0)^2}{4\pi})
\end{displaymath}
\begin{equation}
=u_{\rm c} (\frac{\gamma}{\gamma-1} p_{\rm i}+\frac12\rho_{\rm i} u_{\rm c}^2+\pi_{\rm i}+\frac{B^2}{4\pi})
-\int_0^{x}S_{\rm i}{\rm d}x'
\end{equation}

\begin{displaymath}
\dot{E_{\rm e}}=u^0_{\rm c} (\frac{\gamma}{\gamma-1} p^0_{\rm e}+\frac12\rho^0_{\rm e} (u^0_{\rm c})^2)
\end{displaymath}
\begin{equation}
=u_{\rm c} (\frac{\gamma}{\gamma-1} p_{\rm e}+\frac12\rho_{\rm e} u_{\rm c}^2+\pi_{\rm e})
-\int_0^{x}S_{\rm e}{\rm d}x'
\end{equation}
where $S_{\rm n}$, $S_{\rm i}$, and $S_{\rm e}$ stand for the
source terms in the right hand side of the conservative form of the
total energy equations \ref{hydro1}-\ref{hydro2}. We define $\Lambda=S_{\rm n}+S_{\rm i}+S_{\rm e}$
and we note that $\Lambda=\Lambda_{\rm n}+\Lambda_{\rm i}+\Lambda_{\rm e}$.

 Magnetic flux~:
\begin{equation}
\dot{B}=u^0_{\rm c} B^0=u_{\rm c} B
\end{equation}
 
Integrals involve the source terms describing collisional, chemical,
and thermal exchanges between different fluids, and radiative
losses. The other terms describe the conservative phenomena that share
mass, momentum, and energy between their available reservoirs
(thermal, kinetic, viscous, magnetic...).
 
  In each sector of a steady J or C shock, we will now get algebraic
  relations between the dominant conserved quantities.  We tackle
  successively the following features, in the order in which a parcel of gas
  entering a magnetised shock would meet them~:
\begin{itemize}
 \item 
 the cold magnetic precursor, in which the ion velocity is
 mainly decelerated, and friction starts to brake the neutrals,
 \item
 the hot magnetic precursor, in which the friction has brought the
 temperature to a sufficiently high level that H$_2$ cooling starts to
 play a dominant role (usually above neutral temperatures of 10$^3$~K),
 \item
 the adiabatic front, in which viscosity in the neutrals converts
 their remaining kinetic energy into heat,
\item
 the relaxation layer, in which the gas cools down, is compressed, and
 gets back to a thermal and chemical equilibrium.
\end{itemize}
 Finally, we derive analytical properties of the atomic plateau that follows
 dissociative shock fronts.

\subsubsection{Cold magnetic precursor}

  We assume here and in all the following that the ionisation fraction
  is very low~:

\begin{equation}
\dot{M}=\dot{M_{\rm n}}+\dot{M_{\rm c}}=\rho_{\rm n} u_{\rm n}
\end{equation}
 
  Viscous, ram and thermal pressure of the charges also are negligible
  compared to neutrals and magnetic pressure. Furthermore, since this
  region is far from the adiabatic shock front, neutral viscosity can
  safely be neglected as well.  The total momentum flux is then~:
\begin{equation}
\dot{P}=\dot{P_{\rm n}}+\dot{P_{\rm c}}=\rho_{\rm n} u_{\rm n}^2+p_{\rm n}+\frac{B^2}{8\pi}
\end{equation}

  If the temperature of the neutrals stays low, radiative losses
can be neglected, and the total energy flux is~:
\begin{equation}
\dot{E}=\dot{E_{\rm n}}+\dot{E_{\rm c}}=
u_{\rm n}(\frac{\gamma}{\gamma-1}p_{\rm n}+\frac12\rho_{\rm n} u_{\rm n}^2)+u_{\rm c} \frac{B^2}{4\pi}
\end{equation}

  Finally, conservation of the magnetic flux through the steady region gives~:
\begin{equation}
\dot{B}=B u_{\rm c}  
\end{equation}

We thus get 4 equations with 5 unknowns $\rho_{\rm n}$, $u_{\rm n}$, $p_{\rm n}$, $B$,
and $u_{\rm c}$. One variable can then be chosen to get expressions for all
the others. For example, $p_{\rm n}$ is solution of a quadratic whose coefficients
depend on the shock parameters and the neutral density $\rho_{\rm n}$~:
\begin{displaymath}
p_{\rm n}^2 \frac{2\pi}{\dot{B}^2}\frac{\dot{M}^2}{\rho_{\rm n}^2}\frac{\gamma^2}{(\gamma-1)^2}
\end{displaymath}
\begin{displaymath}
+p_{\rm n}[1-\frac{2\pi \dot{M}}{\dot{B}^2 \rho_{\rm n}}\frac{2\gamma}{\gamma-1}(\dot{E}-\frac12\frac{\dot{M}^3}{\rho_{\rm n}^2})]
\end{displaymath}
\begin{equation}
-\dot{P}+\frac{\dot{M}^2}{\rho_{\rm n}}
+\frac{2\pi}{\dot{B}^2}(\dot{E}-\frac12\frac{\dot{M}^3}{\rho_{\rm n}^2})^2=0
\end{equation} 

As another example, $u_{\rm n}$ is solution of a quadratic whose coefficients
depend on the shock parameters and $u_{\rm c}$~:
\begin{equation}
\label{unuc1}
\frac{\gamma+1}{2(\gamma-1)} \dot{M} u_{\rm n}^2
-\frac{\gamma}{\gamma-1}(\dot{P}-\frac{\dot{B}^2}{8\pi u_{\rm c}^2})u_{\rm n}
+\dot{E}-\frac{\dot{B}^2}{4\pi u_{\rm c}}
=0
\end{equation}

  These relations hold up to the point above which radiative losses
  cannot be neglected anymore (they would
  hold as well in C-shocks upstream this point). In our simulations
  (Paper~I), this corresponds to the point where $T_{\rm n}>10^3$~K. The
  gas then enters the hot magnetic precursor, where H$_2$
  cooling becomes dominant.

\subsubsection{Hot magnetic precursor}

  In this part of the magnetic precursor, ram pressure is directly
transferred into radiation via friction. There is no more increase
in the thermal pressure. Therefore, the neutral thermal pressure
becomes very quickly negligible against magnetic
pressure. It also remains negligible against ram pressure in the rest
of the magnetic precursor. 
  This leads to the following reduced set of equations~:
\begin{equation}
\begin{array}{lcl}
\dot{M}&=&\rho_{\rm n} u_{\rm n}\\
\dot{P}&=&\rho_{\rm n} u_{\rm n}^2+\frac{B^2}{8\pi}\\
\dot{B}&=&Bu_{\rm c}
\end{array}
\end{equation}

 We can derive from this a relation between the speeds of neutrals and
charges that is valid in magnetic precursors (or C-shocks) downstream
 the point where $|\int_0^x\Lambda{\rm d}x|$ dominates over 
$u_{\rm n} \frac{\gamma}{\gamma-1} p_{\rm n}$ (usually, when $T_{\rm n}$ is greater than 10$^3$~K)~:
\begin{equation}
\label{unuc2}
u_{\rm n}=\frac1{\dot{M}}(\dot{P}-\frac{\dot{B}^2}{8\pi u_{\rm c}^2})
\end{equation}
  This last equation is complementary to equation \ref{unuc1}.
   These equations provide a powerful way to test if
  the magnetic field compression is correctly treated in a
  multifluid code. 

\subsubsection{Adiabatic shock front}

In the shock front, the viscous pressure $\pi_{\rm n}$ is one additional
unknown.  We will therefore assume that we know the magnetic field
$B_{\rm p}$ at the end of the magnetic precursor, as well as the amount of
energy radiated away in the precursor $\dot{E}_{\rm p}=\int_{\mbox{\small
precursor}}\Lambda{\rm d}x'$.  Since the shock front is very
tenuous, it is fair to assume that neither the magnetic field nor
the integrated radiative losses will vary across it.

  We can then define the new conserved fluxes for this region~:
\begin{equation}
\begin{array}{lcl}
\dot{P}'&=&\dot{P}-\frac{B_{\rm p}^2}{8\pi}\\
\dot{E}'&=&\dot{E}+\dot{E}_{\rm p}
\end{array}
\end{equation}

  Four equations then combine together~:
\begin{equation}
\begin{array}{lcl}
\dot{M}&=&\rho_{\rm n} u_{\rm n}\\
\dot{P}'&=&\rho_{\rm n} u_{\rm n}^2+p_{\rm n}+\pi_{\rm n}\\
\dot{E}'&=&u_{\rm n}(\frac{\gamma}{\gamma-1}p_{\rm n}+\frac12\rho_{\rm n} u_{\rm n}^2+\pi_{\rm n})+u_{\rm c} \frac{B_{\rm p}^2}{8\pi}\\
\dot{B}&=&u_{\rm c} B_{\rm p}\\
\label{adequ}
\end{array}
\end{equation}
 
  Here, we explicitly deduce the pressure in terms of the density~:
\begin{equation}
p_{\rm n}=
(\gamma-1)(\frac12\frac{\dot{M}^2}{\rho_{\rm n}}-\dot{P}'+
\frac{\dot{E}'-\dot{B}B_{\rm p}/8\pi}{\dot{M}}\rho_{\rm n})
\end{equation}

  We are not aware of any previous analytic expression relating
  pressure to density throughout an adiabatic shock front, even in the
  absence of magnetic fields. This relation is useful to test a code
  in a shock front.

In addition, the post-shock velocity $u_{\rm n}$ 
can be calculated by setting $\pi_{\rm n} = 0$ in equations \ref{adequ}, which gives
the following quadratic equation~:
\begin{equation}
\dot{M}\frac{\gamma+1}{2}u_{\rm n}^2 - \gamma\dot{P}' u_{\rm n} + (\gamma-1)(\dot{E}'-\dot{B}B_{\rm p}/8\pi) = 0
\label{unequ}
\end{equation}
Without magnetic field and energy losses, this quadratic gives the post-shock
velocity and hence the usual compression factor in an adiabatic shock. 
The same kind of reasoning will also provide us with analytic
predictions in dissociative shock fronts (see \ref{derivationplateau}).

\subsubsection{Relaxation layer}
  Here, radiative losses are not negligible anymore, and the equation
  of conservation of energy is left aside. But outside the shock
  front, viscous pressures are negligible, so we only need to make an
  assumption about the magnetic field.  Since we neglect the thermal
  and ram pressure of the charges, the momentum conservation of the
  charges yields~: 
\begin{equation}
\frac{B^2}{8\pi}=\dot{P_{\rm c}}+\int_0^{x}F_{\rm n\rightarrow c}{\rm d}x'
\end{equation}

For low shock velocities, the last integral is dominated by the
magnetic precursor, where most of the ion
deceleration occurs, and $B=B_{\rm p}$ is also a correct
approximation in the relaxation
layer. We are then left with two equations~:
\begin{equation}
\begin{array}{lcl}
\dot{M}&=&\rho_{\rm n} u_{\rm n}\\
\dot{P}'&=&\rho_{\rm n} u_{\rm n}^2+p_{\rm n}\\
\end{array}
\end{equation}
$p_{\rm n}(\rho_{\rm n})$ follows easily~:
\begin{equation}
p_{\rm n}=\dot{P}'-\frac{\dot{M}^2}{\rho_{\rm n}}
\end{equation}
  Note that the intersection of this relation with the thermal
  equilibrium relation $p_{\rm n}(\rho_{\rm n})$ gives the final steady
  post-shock conditions. Similarly, the intersection of
  the algebraic equations in two adjacent sectors of the shock gives
  the physical conditions at the transition
  between the two regions. 

In cases of high shock velocities (greater than 30~kms$^{-1}$ for
the models in Paper~I), a recoupling of the velocities
of charges and neutrals may happen in the relaxation layer, which
builds up an additional magnetic pressure. 
After the recoupling zone, we can assume that $u_{\rm n}=u_{\rm c}$~:
\begin{equation}
\begin{array}{lcl}
\dot{M}&=&\rho_{\rm n} u_{\rm n}\\
\dot{P}&=&\rho_{\rm n} u_{\rm n}^2+p_{\rm n}+\frac{B^2}{8\pi}\\
\dot{B}&=&Bu_{\rm n}
\end{array}
\end{equation}
 $p_{\rm n}(\rho_{\rm n})$ follows easily~: 
\begin{equation}
p_{\rm n}=\dot{P}-\frac{\dot{M}^2}{\rho_{\rm n}}-\frac{\dot{B}^2}{8 \pi
\dot{M}^2}\rho_{\rm n}^2 
\end{equation}

 At the high density end of the relaxation layer, the total pressure
 is dominated by magnetic pressure. This prevents the gas to be
 compressed while it cools down, and leads to much lower compression
 factors. The final state of the gas in this case corresponds to the
 steady isothermal compression factor calculated in the case of
 magnetic field coupled to the gas \citep[see][ equation 2.19a]{DM93}.

 \subsection{Adiabatic dissociative shock}

  In an adiabatic dissociative shock, energy losses can still be
  accounted for by conserved quantities since thermal energy of the gas
  is used to dissociate H$_2$, i.e. converted into internal energy.
We then have~:
\begin{equation}
\begin{array}{lcl}
\dot{E}&=&u_{\rm n}(\frac{\gamma}{\gamma-1}p_{\rm n}+\frac12\rho_{\rm n} u_{\rm n}^2+\pi_{\rm n}) + 
\int_0^{x}R^{\rm d}_{{\rm H}_2}Q {\rm d}x'
\end{array}
\end{equation}
where $Q$ = 4.48~eV is the binding energy of the H$_2$ molecule and
$R^{\rm d}_{{\rm H}_2}$ is the local dissociation rate. The integral term is
related to the change in the flux of H$_2$ molecules by~:
\begin{equation}
 \int_0^{x}R^{\rm d}_{{\rm H}_2}Q {\rm d}x' = Q (0.5 -
 f(x))\frac{\dot{M}x_{\rm H}}{\mu_{\rm H}}
\end{equation}
where $f$ is the fractional abundance of H$_2$ by number relative to
$n_{\rm H}$ (initially $f$ = 0.5), $x_{\rm H}$ is the (fixed) elemental hydrogen
mass fraction, and $\mu_{\rm H}$ is the mass of one hydrogen atom.

  For more clarity, we neglect here the effects of the charged fluid,
 but it would be straightforward to include the magnetic pressure and
  radiative effects using primed shock parameters like in the previous
  subsections.  We then obtain a modified version of
equations \ref{adequ}, including the H$_2$ dissociation energy~:
\begin{equation}
\begin{array}{lcl}
\dot{M}&=&\rho_{\rm n} u_{\rm n}\\ 
\dot{P}&=&\rho_{\rm n} u_{\rm n}^2+p_{\rm n}+\pi_{\rm n}\\
\dot{E}&=&u_{\rm n}(\frac{\gamma}{\gamma-1}p_{\rm n}+\frac12\rho_{\rm n} u_{\rm n}^2+\pi_{\rm n})
+(1-2f) \frac{\dot{M}Qx_{\rm H}}{2\mu_{\rm H}}
\end{array}
\end{equation}
 
  Unfortunately, we do not have a fixed relation between $p_{\rm n}$ and
$\rho_{\rm n}$ as in the non-dissociative case, because the H$_2$ fraction
$f$ varies across the front and adds to the unknowns. 
But the conditions at the end of the front can be found if one sets
$\pi_{\rm n}=0$ and $f=f^*$ where $f^*$ is the H$_2$ fraction at the end
of the shock. So doing, we get an equation similar to relation \ref{unequ}~:
\begin{equation}
\dot{M}\frac{\gamma+1}{2}u_{\rm n}^2 - \gamma\dot{P} u_{\rm n} + 
(\gamma-1)(\dot{E}-(1-2f^*) \frac{\dot{M}Qx_{\rm H}}{2\mu_{\rm H}}) = 0
\end{equation}
 
 We simplify this equation by assuming the high Mach number regime,
for which $\dot{P}=u^0 \dot{M}$ and $\dot{E}=\frac12(u^0)^2\dot{M}$~:
\begin{equation}
\frac{\gamma+1}{2}u_{\rm n}^2- \gamma u_{\rm n} u^0 +\alpha\frac{\gamma-1}2(u^0)^2=0
\label{compequ}
\end{equation}
  where we defined $\alpha=1-(1-2f^*) \frac{Qx_{\rm H}}{(u^0)^2\mu_{\rm H}}$.
  $1-\alpha$ measures the relative decrease in energy flux through the
  shock front due to H$_2$ dissociation.  In a non-dissociative shock,
  $\alpha=1$ since $f^*=0.5$. $\alpha$ reaches a minimum of $0.74$ for
  a fully dissociative shock ($f^*=0$) just above the dissociation
  limit $u^0 = u_{\rm d} \simeq 30$~kms$^{-1}$, and tends to
  one again for very high shock speeds (where the H$_2$ dissociation
  energy becomes negligible compared to the kinetic flux).

  The compression factor through such a shock can now be computed~:
\begin{equation}
 C=\frac{u^0}{u_{\rm n}}=\frac{\rho_{\rm n}}{\rho^0}=
\frac{\gamma+\sqrt{\gamma^2(1-\alpha)+\alpha}}{(\gamma-1)\alpha}
\end{equation}
  The usual compression $C_{\rm a}=\frac{\gamma+1}{\gamma-1}$ is recovered
  when $\alpha=1$. $C=5.8$ for $\gamma=5/3$ and $\alpha=0.74$.

 We can get a simple expression for the  temperature of
 the atomic plateau ($f^*=0$) if we neglect the post-shock ram pressure
 (so that $p_{\rm n}=\rho^0(u^0)^2$). This is only 20\% accurate for the
 compression factor obtained (we use $\gamma=5/3$)~:
\begin{equation}
T_{\rm p}=\frac{\mu_{\rm p}}{{\rm k}_{\rm B}}(u^0)^2 \alpha
\label{anaplateau}
\end{equation}
where $\mu_{\rm p}$ is the mean molecular weight in the plateau.  $T_{\rm p}$ is
hence nearly quadratic in the entrance velocity for strong shocks. 

The
knowledge of the compression factor, along with the assumption of
steadiness of the adiabatic front provides us with the velocity of the
front relative to the piston in the adiabatic phase of a dissociative
shock front~:
\begin{equation}
v=u/(C-1)
\label{velodiss}
\end{equation}

\label{derivationplateau}

\subsection{Validation of the analytical results on examples}

We verified the analytical relations derived above by comparing
with numerical magnetic shock simulations from Paper~I.

First, we find that the temperature and density of the atomic plateau
are well predicted by the formulae \ref{anaplateau} in most
dissociative shocks. This was expected, since the width of the
adiabatic shock is so small that it is very likely to be in a steady
state~: the sound crossing time of this feature is much shorter than
the time of variation of the entrance shock speed. In addition, in
strongly oscillating shocks (weakly dissociative case of Paper~I), the
maximum expansion of the front, corresponding to the adiabatic, fully
dissociative phase, coincides with the velocity given by equation
\ref{velodiss} (see Fig. 4d of Paper~I).

\begin{figure}[h]
\centerline{\psfig{file=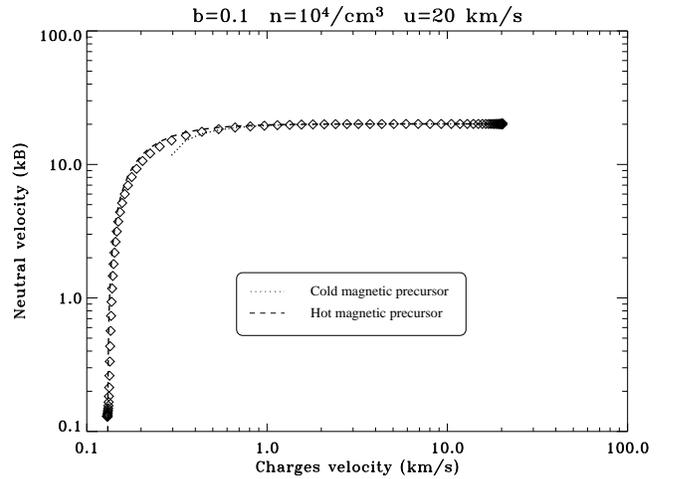,angle=-90,width=8.8cm}}
\caption{Steady state analytic relations between the
velocities of neutrals and charges are compared to an overlaid steady C-shock
(diamonds). The parameters of the shock are $u=20$~kms$^{-1}$,
$n=10^4~$cm$^{-3}$ and $b=0.1$, time is $t=10^5$~yr. The velocities in the shock frame are computed using
a velocity of the shock front of 0.13 kms$^{-1}$, inferred from
Fig. 3d of Paper~I.}
\label{checku}
\end{figure}

We also verified the relations predicted for non-dissociative shocks.
As an example, in Fig. \ref{checku}, we check the relations
\ref{unuc1} and \ref{unuc2} against the final steady-state of a
C-type shock (diamonds). The agreement is very good, confirming that
magnetic compression is correctly treated in the code.

\begin{figure}[h]
\centerline{\psfig{file=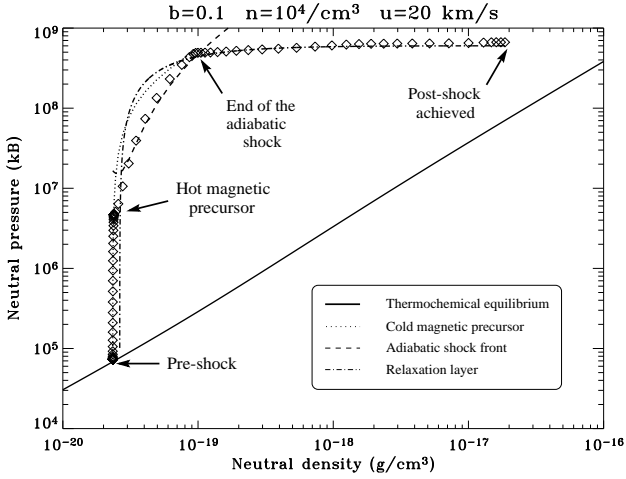,angle=-90,width=8.8cm}}
\caption{Steady state analytic relations between 
pressure and mass density are compared to an overlaid future C-shock
(diamonds). The parameters of the shock are $u=20$~kms$^{-1}$,
$n=10^4~$cm$^{-3}$ and $b=0.1$, time is $t=10^2$~yr. The additional
necessary parameters $B_{\rm p}$ and $\dot{E}_{\rm p}$ are read in the shock model
at the end of the precursor (they are {\it not} fitted).}
\label{checkp}
\end{figure}

Finally, in figure \ref{checkp} we plot in diamonds the state of the
gas for the same shock in a snapshot at $t = 100$ yrs, i.e. well
prior to steady (C-type) state. We overlay the algebraic relations
previously derived for the various shock regions. The agreement turns
out to be very good. This comforts us in the ability of the code to
reproduce the conservation equations. It also points out the fact that
steady equations may well be valid to describe a shock at early times,
even though it has not reached its final steady-state. It is to
address the domain of validity of this ``quasi-steady''
approximation that we set up the technique and
tests described in the next section.

\section{Validity of the quasi-steady assumption 
in time-dependent shocks}
\label{quasisteady}
  In this section we develop a method to characterise the local
  steady velocity for each variable separately, and we use it to
  test the ''steadiness''  of various shock regions in the simulations of
  Paper~I.

\subsection{Local steady velocities and quasi-steady state}

  Consider $y$, one of the $N+6$ state variables that enter the set of
  equations \ref{hydro1}-\ref{hydro2}.  Its evolution equation in the frame of the
  piston can be cast in the following form~:
\begin{equation}
\frac{\partial y}{\partial t}+\frac{\partial (u y)}{\partial x}=s
\end{equation}
where $u= u_{\rm n}$ or $u_{\rm c}$ is the velocity of the fluid associated with
$y$, and $s$ is a term that does not depend on the reference frame.

We define the local steady velocity $v_y$ for variable $y$ as~: the
velocity of a reference frame in which the time derivative in the
evolution equation vanishes. Hence~:
\begin{equation}
\frac{\partial }{\partial x}[(u-v_y) y]=s 
\end{equation}

When $y$ does not involve a velocity, i.e. does not depend on the
reference frame, a direct expression follows for the velocity $v_y$~:
\begin{equation}
v_y=[\frac{\partial (uy)}{\partial x}-s]/\frac{\partial y}{\partial x}.
\label{vsteady}
\end{equation}

  One has to be more careful when $y$ depends on $u$. For
  example, in the case $y=\rho u$, $v_y$ is given implicitly by the
  following quadratic equation~:
\begin{equation}
\label{steadymom}
v_y^2\frac{\partial \rho}{\partial x}
-2v_y\frac{\partial y}{\partial x}
+\frac{\partial (uy)}{\partial x}=s
\end{equation}

 The expression \ref{vsteady} is singular when $\frac{\partial
 y}{\partial x}=0$. This can easily be understood~: if the profile of
 $y$ is flat, any velocity will do.  It is therefore crucial to take
 into account the finite numerical precision when trying to evaluate
 $v_y$ with these expressions. Indeed, roundoff errors can make the
 gradient of $y$ non zero even in places where it should be.

  Finally, expressions \ref{vsteady} and \ref{steadymom} yield a way
  of characterising the ``steadiness'' of the flow. At a given
  position $x$, if $v_y$ does not depend on $y$, then there is indeed
  a frame moving at a velocity $v(x)$ in which none of the variables
  is changing in time.  Furthermore, if this velocity is constant
  throughout an extended region, then this whole region is moving
  ``en-bloc'' at velocity $v$ and can be modelled with a truncated
  steady-state model. We say that this region is in a {\it
  quasi-steady state}.

\subsection{Validity of the quasi-steady state in time-dependent shocks}
  For a selection of time steps of each of the dynamical models that
  we simulated in Paper~I, we computed the steady
  velocities $v_y$ in each zone for each variable $y$.

  We evaluated the numerical noise in the following way~: we computed
  the change $\delta v_{yy'}$ in the steady velocity $v_{y}$ when each
  variable $y'$ was changed by 10$^{-4}$ in relative value
  (corresponding to our guess for the numerical precision). We then
  estimated the numerical noise $\sigma_y$ on
  variable $y$ by~:
\begin{equation}
\sigma_y^2=\sum_{y'}(\delta v_{yy'})^2
\end{equation}
  The noise-weighted mean steady velocity over all
variables of a subset $S$ was then computed in each zone, as well as the
corresponding numerical noise $\sigma_{\rm noise}$~:
\begin{equation}
v =(\sum_{y\in S}v_y/\sigma_y^2)/(\sum_{y\in S}1/\sigma_y^2)
\end{equation}
\begin{equation}
\sigma_{\rm noise}^2 = 1/(\sum_{y\in S}1/\sigma_y^2)
\end{equation}
 For charges and neutral momentum, both roots of quadratic
 \ref{steadymom} where included. But for reasons that will become
 clear in the next subsections, magnetic field was excluded from this
 mean. Finally, the variance of individual 
$v_y$ values about  this mean velocity and the corresponding
$\chi^2$ were estimated~:
\begin{equation}
\sigma^2 =[\sum_{y\in S}(v_y-v)^2/\sigma_y^2]/(\sum_{y\in S}1/\sigma_y^2)
\end{equation}
\begin{equation}
\chi^2 = \frac1{\#S}\sigma^2/\sigma_{\rm noise}^2 
= \frac1{\#S}\sum_{y\in S}(v_y-v)^2/\sigma_y^2
\end{equation} 
where $\#S$ is the number of variables in the subset $S$.  

If the
numerical noise is well estimated, a value of $\chi^2 \simeq 1$
indicates that the dispersion of invidual $v_y$ values is consistent
with local numerical noise, i.e. that there may exist a common steady
velocity for all the variables of the subset $S$ at that
position. Regions where this is fulfilled {\it and} $v$ is constant
are the quasi-steady regions for the set $S$.

  A less strict criterion can be chosen for the local steadiness if we
  think of the ratio $v/\sigma$ as a ``signal to noise''~: even if
  $\chi^2$ is high, it may be possible that the ratio $v/\sigma$ is
  high. In this case, the steady velocities corresponding to different
  variables are not equal, but they are close to one another~:
  therefore, a common steady velocity $v$ is a good approximation.

 It turns out that we compute very low values of $\chi^2$ in quite a
 few zones. This indicates that the numerical precision we have in these
zones is far better than our estimate of $10^{-4}$. We hence rather use the
criterion based on the ratio $v/\sigma$ and define the quasi-steady
regions as the regions with a constant velocity $v$ {\it and } a good
``signal to noise'' ratio.

The subset of variables $S$ on which the averages are computed should
be the whole set of variables. However, the last section of this paper
needs only that the dynamically important variables be in a quasi-steady
state. Therefore we present here the results for a subset $S$ including
the temperatures of the three fluids, the four velocities (roots
of equation \ref{steadymom} for $y=\rho_{\rm n} u_{\rm n}$ and $y=\rho_{\rm c} u_{\rm c}$), 
He, H$_2$, H$_2$O, CO and OH densities. We do not include the magnetic field
in magnetised shocks, because its steady velocity differs from the other
variables as we will show in the next section. We also did the calculation
for $S$ including all the variables but the magnetic field, and found
that it did not change the general conclusions : the signal to noise
ratio is slightly less good, and a few zones are not quasi-steady
anymore because of marginal chemical species having a different steady velocity
than the bulk of the variables.

We summarise the results of our investigation in the next two subsections,
devoted respectively to non-magnetised and magnetised shocks.

\subsubsection{J-type shocks ($B$=0)}
\label{vsteadJ}

{\it Non-dissociative J-type shocks}

\begin{figure}[h]
\centerline{\psfig{file=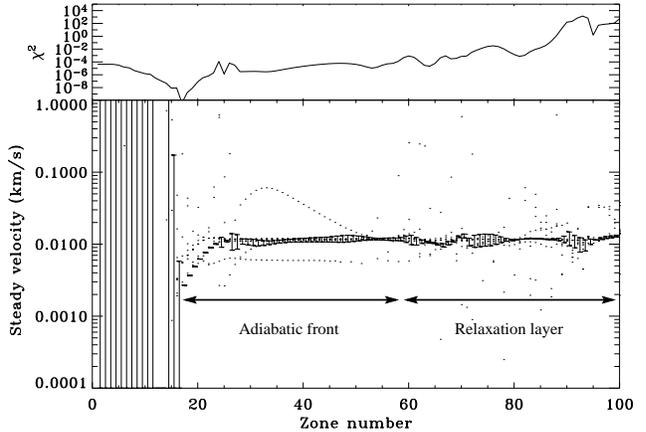,angle=-90,width=8.8cm}}
 \caption{We plot the $\chi^2$ values and steady velocities for each
 variables of $S$ in each zone of snapshot $t=200$ years for the shock with
 parameters $b=0$, $n=10^4~$cm$^{-3}$, and $u=20$~kms$^{-1}$. Each dot
 represents the steady velocity of one variable computed thanks to
 expressions \ref{vsteady} and \ref{steadymom}. The error bars are
 $v\pm \sigma$ evaluated zone per zone on these values. We indicate
 the computational domains associated to the adiabatic front and the
 relaxation layer of the shock.}
\label{vsJ}
\end{figure}

Figure \ref{vsJ} shows the $\chi^2$ as well as $v$ with error bars
$\pm \sigma$ in each zone of a typical snapshot of a non-dissociative
J-type shock. The adiabatic front and relaxation layer show a good
``signal to noise'' ratio and a flat steady velocity. On the
contrary, the pre-shock zones show a huge dispersion around the steady
velocity, expected from the homogeneity of the medium there. The fact
that we retained both velocities from expression \ref{steadymom} does
not alter the results, because the numerical errors on velocities are
actually much larger than for the other variables. 

We find that all of our non-dissociative J-type shocks are in a
quasi-steady state from the adiabatic phase to the steady phase (only
the initial formation of the shock front is not quasi-steady).
Therefore, a snapshot of such a shock will always coincide with the
truncated structure of a J-type steady state. 
 
\begin{figure}[h]
\centerline{\psfig{file=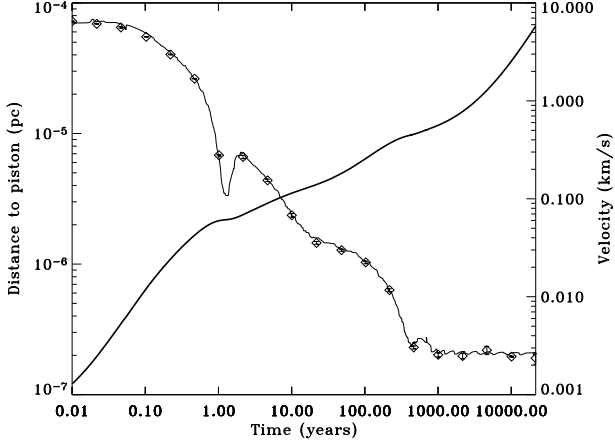,angle=+90,width=8.8cm}}
\caption{Trajectory and velocity away from the piston of the J-shock
 with parameters $b=0$, $n=10^4~$cm$^{-3}$, and $u=20$~kms$^{-1}$ (from
 Paper~I). Overlaid diamonds are the steady velocities $v$ averaged
 over all variables and all zones, for each snapshot
 analysed.}
\label{checkvJ}
\end{figure}

In Paper~I, we plotted the trajectory of the point of maximum ratio of
viscous over thermal pressure (see Fig. 1b). We compute here
the average steady velocity $v$ over the whole structure of the shock
(trimmed from the pre-shock values) at various times and overlay it
over this trajectory in Fig. \ref{checkvJ}. Error bars show the good
consistency of the test, and the correspondence of $v$ with the
velocity at which the viscous shock front moves away from the piston.

 Note that the entrance velocity in the shock is not the upstream
 velocity $u$ of the fluid towards the piston, but rather $u^0 = u+v$.
 Hence, the entrance shock speed for the truncated steady shock is
 evolving in time. In Sect. \ref{construction}, we present a way to
 reconstruct this evolution.
 
 \medskip
{\it Dissociative J-type shocks}

\begin{figure}[h]
\centerline{\psfig{file=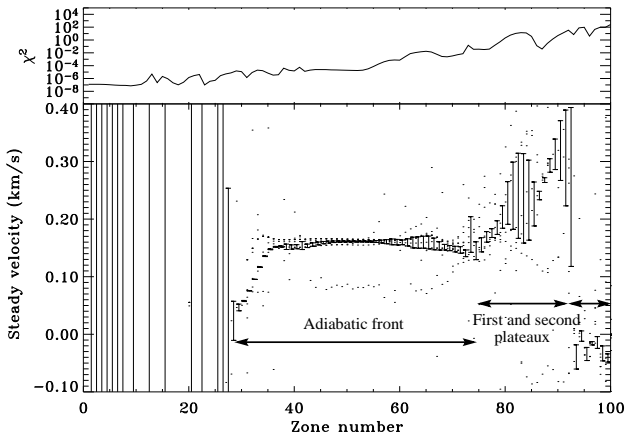,angle=-90,width=8.8cm}}
\caption{Same as Fig. \ref{vsJ}, but for a partly ionising
shock of parameters $b=0$, $n=10^4~$cm$^{-3}$, and $u=40$~kms$^{-1}$ at time
$t=220$ years. Here, the scale of the plot is {\it linear}, so
that the dispersion is in fact much greater than for the
non-dissociative shocks, even in the second plateau.  }
\label{vsJd}
\end{figure}

  On the contrary, dissociative shocks are almost never in a
  quasi-steady state. For weakly dissociative velocities, we could not
  come up with a coherent picture. This was expected, since
  these shocks are highly unstable with large bouncing oscillations
  between a fully dissociative expansion phase and a non-dissociative
  recoil phase (Paper~I). Partly ionised shocks are less
  unstable. We illustrate their behaviour with a typical example
  shown in Fig. \ref{vsJd}.  The adiabatic front is generally in a
  quasi-steady state with velocity roughly equal to the velocity of
  the viscous maximum, but the dispersion is much higher
 than for non-dissociative shocks.  The
  first plateau that follows (with H ionisation and Lyman cooling) is not
  at all in a quasi-steady state~: the dispersion is huge and the mean
  velocity is not even constant. The second plateau (H
  recombination) seems rather quasi-steady, but with a velocity much
  lower than the adiabatic shock front. 
   This warns us that steady-state diagnostics may be
  hopeless for weakly non-dissociative shocks, and that we have to be
  cautious for partly ionising shocks.

\subsection{CJ-type and C-type shocks}
\label{vsteadC}
\begin{figure}[h]
\centerline{\psfig{file=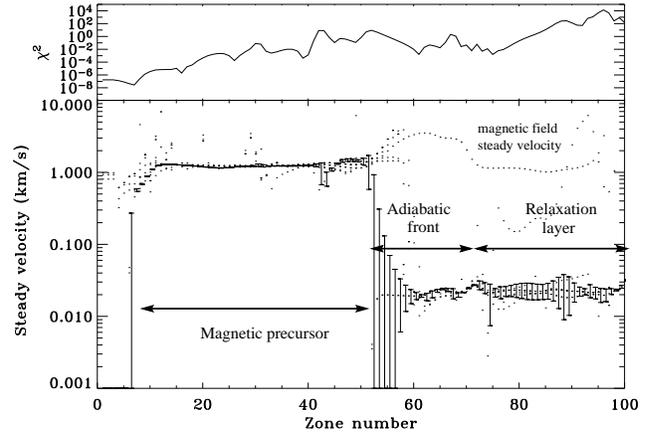,angle=-90,width=8.8cm}}
\caption{Same as Fig. \ref{vsJ}, but for time $t=100$ years of the
 future C-shock with $b=0.1$, $n=10^4~$cm$^{-3}$, and $u=20$~kms$^{-1}$.
 We also show the steady velocity for the magnetic field.  }
\label{vsC}
\end{figure}

Figure \ref{vsC} shows $v$ with error bars $\pm \sigma$ in each zone
of an early snapshot of a non-dissociative C-type shock.  The shock
then has a composite CJ structure made of a magnetic precursor,
followed by a non-dissociative adiabatic front and a relaxation
layer. We find that the magnetic precursor is in a quasi-steady
state with a high steady velocity. This is in agreement with the
remark of \citet{CPF98}. At the very end of the magnetic precursor,
the large dispersion in steady velocities is due to the fact that two
different steady velocities coexist among the variables.  Following
this, the adiabatic front and relaxation layer appear to be in a
quasi-steady state, but with a much lower velocity than the magnetic
precursor.  However, strictly speaking, it is not a real quasi-steady
state, as the steady velocity for the magnetic field remains
equal to the high steady velocity of the magnetic precursor.

\begin{figure}[h]
\centerline{\psfig{file=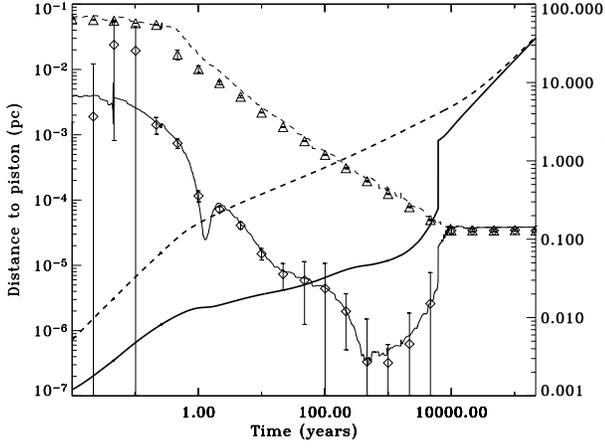,angle=+90,width=8.8cm}}
\caption{Same as Fig. \ref{checkvJ}, but for a C-shock of parameters
 $b=0.1$, $n=10^4~$cm$^{-3}$ and $u=20$~kms$^{-1}$. Curves plot the
 trajectory and velocity of the neutral (solid) and charged (dashed)
 viscous fronts. Diamonds are the quasi-steady velocities of the
 relaxation layer and adiabatic front. Triangles are the quasi-steady
 velocities of the magnetic precursor.
}
 \label{checkvC}
\end{figure}

 Figure \ref{checkvC} shows that the two steady
 velocities (relaxation layer and precursor) correspond well to
 the velocities of the viscous maxima of neutrals and
 charges determined by time-derivation of their trajectory.  As time
 evolves, the velocity of the magnetic precursor and that of the relaxation
 layer get closer to one another, and finally coincide after the
 J-front has disappeared. The C-type structure is then in a
 quasi-steady state as a whole. 

In principle, one should then be able to model an early age of a low
velocity C shock by combining a truncated C-type model with a
truncated J-type model (in which the magnetic field is treated
appropriately).  The problem is a bit more complex than in the J-type
case because we need now to determine two different truncation
distances and two sets of entrance parameters, but we will show in
Sect. \ref{buildC} that it is possible to solve.

 This picture holds for all shocks with magnetic
 field, as long as there is no dissociation or ionisation plateau. In
 the case of dissociative velocities, the same problems described in
 the previous subsection arise in the corresponding features of the
 relaxation layer.

This picture is the same for all magnetic shocks. The only difference
between CJ-type and C-type is whether or not the J-front has
disappeared when the steady velocities of the relaxation layer and the
magnetic precursor converge. From this remark, we will obtain in 
Sect. \ref{CJstate} a way to assess if a low velocity
shock will eventually become a steady CJ-type shock.

\section{Time-dependent constructions of shocks at early times}
\label{construction}
Here, we derive methods of reconstruction of time-dependent shocks
using truncated steady models. Those constructions will be meaningful
only for the shocks in which the quasi-steady state has been validated
at all times, although they can in principle be realised in any shock.
Due to their different complexities, we treat successively the case of
non-dissociative J-type shocks and non-dissociative magnetised shocks.

\subsection{Non-dissociative J-type shocks}
\label{buildJ}

 Section \ref{vsteadJ} has shown that the whole structure of
 non-dissociative J-type shocks is at all times quasi-steady. 
One may then safely fit truncated
 steady models to observations. The fitted parameters would be the
 entrance parameters in the shock frame and the truncation
 distance. But one would then like a method to relate these
 parameters to the parameters of the shock in the
 piston frame, and to the age of the shock.
 Conversely, one would like to build at will a snapshot of a shock 
 of given age and parameters (in the piston frame), with the only help
 of a steady state code. We come up with such a procedure in the
 following.

  A steady-state code provides us with the steady profile of any
  variable in the frame of the shock front for a given set of
  entrance parameters $(u^0,n)$. Say, the
  steady velocity $u_{\rm s}(u^0,n;x)$, where $x$ is the distance from the
  shock front. If we are given an inflow speed and density 
   $(u,n)$ in the frame of the piston and a time $t$,
  the problem is to find what is the entrance velocity $u^0$ at the same time
  $t$ in the frame of the shock front as well as the corresponding
  distance $r$ between the shock front and the piston.

\begin{figure}[h]
\centerline{\psfig{file=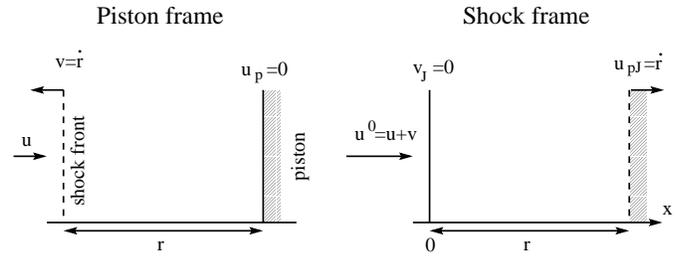,angle=-90,width=8.8cm}}
\caption{Schematical view of a J-type shock in the piston frame and in the 
shock frame.
}
\label{sketchJ}
\end{figure}

To help set up the notations in both frames of the shock
and of the piston, we sketched in Fig. \ref{sketchJ} the different
lengths and velocities involved. 

 At any given time, velocities in the shock frame are found by
adding $v= \dot{r}$ to velocities in the piston frame. The entrance
shock speed is then~:
\begin{equation}
u^0=u+\dot{r},
\end{equation}
while the velocity at the piston --- which must be null in the piston frame ---
is, in the J-shock frame~:
\begin{equation}
u_{\rm s}(u^0,n;r)=\dot{r}.
\end{equation}
These relations combine to give an implicit equation linking $\dot{r}$ to $r$~:
\begin{equation}
\dot{r}=u_{\rm s}(u+\dot{r},n;r)
\label{r}
\end{equation}

Furthermore, in a quasi-steady state, mass conservation requires that
$u_{\rm s}(u^0,n;r)\times C = u^0$, i.e. $\dot{r} = v = u/(C-1)$ where $C$ is the
compression factor at the piston.  From the adiabatic phase ($C =
4$) to the steady state ($C << 1$), the speed $v$ of the front
thus decreases from $\frac13u$ to nearly $0$.  Therefore, if
the steady state code provides $u_{\rm s}(u^0,n;r)$ for a range of
velocities $u<u^0<\frac43u$, the equation \ref{r} is an implicit
ordinary differential equation straightforward to integrate up to time
$t$ with initial conditions $r=0$ and $\dot{r}=\frac13u$. Actually, an
interpolation between a few steady models might be sufficient to get
accurate results.

  Conversely, if the problem is to recover the time from a
  steady model truncated at distance $R$, we only have to integrate
  equation \ref{r} backward in time up to the point where $r=0$ and
  compute $t=\int_0^{R}\frac1{\dot{r}} {\rm d}r$.

\medskip
{\it High compression factor approximation~:}

 For high compression factors, the final $\dot{r}$ is small
 enough to be neglected with respect to $u$, which makes the
 computation even easier. The integral yielding the age is dominated
 by the very low velocities, which are also the most recent
 ones. Therefore, we only need one steady state model
 $u_{\rm s}(x)=u_{\rm s}(u,n,B;x)$. The age of such a truncated shock is
 simplified in the following way.
\begin{equation}
t=\int_0^{R}\frac1{u_{\rm s}}{\rm d}x
\end{equation}
  In this last expression, one recognises the flow time across the shock.
  Since for strongly radiative shocks, the compression factor rises
  very quickly, this approximation is valid even for very young
  ages. For example, the shock of parameters $b=0$,
  $n=10^4$~cm$^{-3}$, and $u=20$~kms$^{-1}$ has a compression factor of 10
  at as early as $t=1$ year (see Fig. \ref{checkvJ}).

\subsection{Non-dissociative magnetised shocks}
\label{buildC}
\begin{figure}[h]
\centerline{\psfig{file=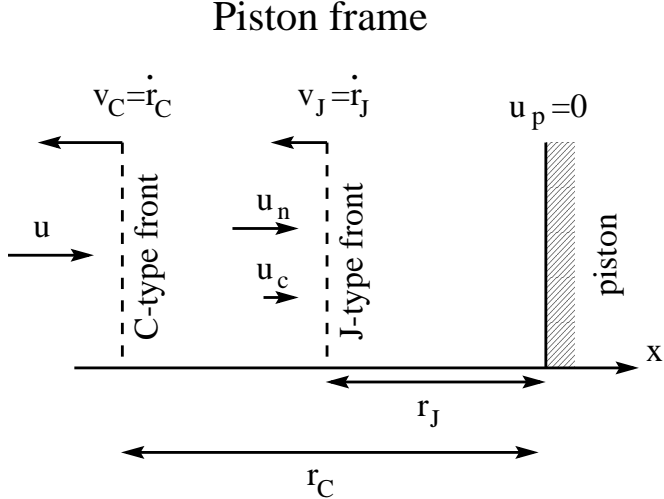,angle=-90,width=8.8cm}}
\caption{Schematical view of an early magnetised shock in the piston frame.}
\label{sketchC}
\end{figure}

  The  analysis of Sect. \ref{quasisteady} showed that
  low-velocity magnetised shocks are composed of two quasi-steady
  regions~: a magnetic precursor, and a non dissociative J-type
  feature. In faster shocks,  where the entrance
  velocity in the J-type feature is dissociative, the
  J-type structure is not in a quasi-steady state, although the
  magnetic precursor is. We thus restrict our analysis to the
  non-dissociative cases.

  In principle, one should then be able to model a snapshot of such a
  shock by gluing together two truncated steady C and J models.
  The problem is to determine the entrance parameters and lengths of
  each of the two shock features, for a given time $t$ and a given set
  of parameters $(u,n,B)$ in the piston frame. A rigorous construction
  method is outlined below.

 In the following, $r_{\rm C}$ and $r_{\rm J}$ denote respectively the distance of
 the fronts of the C-type and the J-type features with respect to the
 piston (see Fig. \ref{sketchC}). Variables computed in the
 reference frame of the C-type feature are denoted with C
 subscripts. Variables in the J-type feature frame are specified with
 J subscripts. Entrance parameters in the shocks have a 0
 superscript.
 
 Now, let us assume that we know the positions $r_{\rm C}$ and
 $r_{\rm J}$ at current time $t$.  To solve for their evolution, we need
 to find equations that will determine $\dot{r}_{\rm C}$ and $\dot{r}_{\rm J}$.

The entrance velocities in the C-type feature are simply:
\begin{equation}
\label{entranceC}
\begin{array}{lcl}
u_{\rm nC}^0&=&\dot{r}_{\rm C}+u\\
u_{\rm cC}^0&=&\dot{r}_{\rm C}+u\\
n^0_{\rm C}&=&n\\
B^0_{\rm C}&=&B
\end{array}
\end{equation}
The steady-state code for the C-type feature then provides us with the
entrance values of velocities, densities, and magnetic field at the
position of the J-front, $x = r_{\rm C}-r_{\rm J}$. After a suitable change of
reference frame for the velocities, they determine the entrance
parameters in the J-type shock~:
\begin{equation}
\label{entrance}
\begin{array}{lcl}
u_{\rm nJ}^0&=&\dot{r}_{\rm J}-\dot{r}_{\rm C}+u_{\rm nC}(u+\dot{r}_{\rm C},n,B;r_{\rm C}-r_{\rm J})\\
u_{\rm cJ}^0&=&\dot{r}_{\rm J}-\dot{r}_{\rm C}+u_{\rm cC}(u+\dot{r}_{\rm C},n,B;r_{\rm C}-r_{\rm J})\\
n_{\rm nJ}^0&=&n_{\rm nC}(u+\dot{r}_{\rm C},n,B;r_{\rm C}-r_{\rm J})\\
n_{\rm cJ}^0&=&n_{\rm cC}(u+\dot{r}_{\rm C},n,B;r_{\rm C}-r_{\rm J})\\
B_{\rm J}^0&=&B_{\rm C}(u+\dot{r}_{\rm C},n,B;r_{\rm C}-r_{\rm J})
\end{array}
\end{equation}

 The steady-state J-shock must then be integrated. A multifluid
treatment is necessary, since $u_{\rm n}$ and $u_{\rm c}$ at the entrance of the
J-front are different. Furthermore, a special treatment
of magnetic field compression is necessary, since Sect.
\ref{vsteadC} showed that the steady velocity for the magnetic field
in the J-type feature is not $\dot{r}_{\rm J}$, but remains
 the same as in the magnetic precursor, namely $v_B =
\dot{r}_{\rm C}$.  It means that the product of $B$ with the
velocity of charges {\it computed in the frame of the C shock} remains
constant through the J-type feature~:
\begin{equation}
\label{magJshock}
B \times (u_{\rm cJ}-\dot{r}_{\rm J}+\dot{r}_{\rm C}) = B_{\rm C}^0 \times u_{\rm cC}^0
\end{equation}
Therefore, the evolution of the J-type feature
depends not only on the entrance parameters determined above but also
on $\dot{r}_{\rm C}-\dot{r}_{\rm J}$.

 As in the non-magnetic case, the derivatives $\dot{r}_{\rm C}$ and
 $\dot{r}_{\rm J}$ are then determined by stating that both charges and
 neutrals have to be at rest near the piston, i.e., in the frame of
 the J-front~:
\begin{equation}
\label{rest}
\begin{array}{lcl}
\dot{r}_{\rm J}&=&
u_{\rm nJ}(u_{\rm nJ}^0,u_{\rm cJ}^0,n_{\rm nJ}^0,n_{\rm cJ}^0,B_{\rm J}^0,\dot{r}_{\rm C}-\dot{r}_{\rm J};r_{\rm J})\\
\dot{r}_{\rm J}&=&
u_{\rm cJ}(u_{\rm nJ}^0,u_{\rm cJ}^0,n_{\rm nJ}^0,n_{\rm cJ}^0,B_{\rm J}^0,\dot{r}_{\rm C}-\dot{r}_{\rm J};r_{\rm J})\\
\end{array}
\end{equation}
We thus get two independent implicit equations for
$\dot{r}_{\rm J}$ and $\dot{r}_{\rm C}$.  Numerical techniques to solve these
equations still need to be designed, but should not be too hungry in
CPU time. 

\subsubsection{Final steady state: C or CJ ?}
	\label{CJstate}
  We have now a means of computing the time evolution of a magnetised
  shock with only a steady-state code. One
  should then be able to integrate it from initial conditions
  $r_{\rm J}=r_{\rm C}=0$ up to the steady state where
  $\dot{r}_{\rm J}$ and $\dot{r}_{\rm C}$ are equal constants.  If during the
  evolution the entrance velocity in the J-shock $u_{\rm nJ}^0$ becomes
  subsonic, then one should stop the integration because the
  J-shock disappears, and the remaining sound wave propagates through
  the structure until a stationary C-type structure is obtained.  If
  $u_{\rm nJ}^0$ stays supersonic when $\dot{r}_{\rm J}$ and $\dot{r}_{\rm C}$ are
  equal constants, then the result is a CJ-type shock steady-state.
 Hence, one is forced to integrate over time $\dot{r_{\rm J}}$ and
 $\dot{r_{\rm C}}$ given implicitly by equations \ref{rest} to know what is
 the final steady-state corresponding to a given set of parameters
 $(u,n,B)$. 

However, one could also think about solving equations
 \ref{entrance} and \ref{rest} for given arbitrary values of the final
 front velocity $v_{\rm f}=\dot{r}_{\rm J}=\dot{r}_{\rm C}$. The result would be a
 series of physically consistent steady CJ-type states, each
 characterised by a different distance $r_{\rm C}-r_{\rm J}$. Only one of these is
 selected by the time evolution but, if the entrance parameters
 $(u,n,B)$ are allowed to evolve in time, it might be possible that
 several (or even all) of these final states can be realised. The
final state would then depend on the evolution history of the entrance
parameters.
  
 In fact, a very easy way to exhibit one of those CJ-type steady-states
would be to use a multifluid steady-state code, and trigger the viscous
dissipation in the neutral at a given position  $r_{\rm C}-r_{\rm J}$ where
the neutral velocity is still supersonic. 

\subsubsection{Low velocity, high compression factor approximations}
  
For all the low velocity cases encountered in our simulations,
we noted that after one year of time, the velocity of the charges was already
almost brought to rest at the end of the magnetic precursor. This
approximation yields the following equation~:
\begin{equation}
\dot{r}_{\rm C}=u_{\rm cC}(u+\dot{r}_{\rm C},n,B;r_{\rm C}-r_{\rm J})
\label{rc}
\end{equation}
which implicitly gives the velocity $\dot{r}_{\rm C}$. The velocity $\dot{r}_{\rm J}$
can then be retrieved by solving the first equation of the set \ref{rest}.

  Just like for the J-type shocks, high magnetic compression factors
  will lead to $\dot{r}_{\rm C}$ negligible before $u$, and will facilitate
  the integration of equation \ref{rc}. In this case, and if in addition
  $r_{\rm J}<<r_{\rm C}$, the age of the shock is given by :
\begin{equation}
t=\int_0^{r_{\rm C}-r_{\rm J}}\frac1{u_{\rm cC}}{\rm d}x
\end{equation}
which is the flow time of the charges across the magnetic precursor.


\section{Discussion}
The analytic relations we found make good benchmarks for testing
codes. In addition, they might provide some theoretical basis for
further investigation of the properties of these shocks in the
parameter space.

  The quasi-steady state analysis of shocks opens a new field of
  possibilities for the steady-state codes. We compare hereafter our
  method to previously used algorithms, and sketch possible extensions
  of our method.

\subsection{Comparison with previous work}

  Our quasi-steady state analysis of J-shocks justifies the use of 
truncated steady-state J-shocks by \citet{R88}. We provide more
theoretical basis to link the true age of the shock to the truncation
distance used.

  \citet{FP99} and \citet{L02} use simple algorithms to produce mixed
C-type and J-type features to mimic time-dependent magnetised shocks.
\citet{L02} greatly improved the method used by \citet{FP99} since
they keep the multifluid treatment of the flow through the relaxation
layer.  They just switch on viscosity in the neutral fluid when they
encounter a sonic point. The present analysis gives a less heuristic
way to know at which point the viscosity should be switched on, and 
Paper~I has already shown that it can be way upstream a sonic point.
Furthermore, we specify that a change of velocity
frame has to be done at the end of the magnetic precursor,
except for the magnetic field equation. Finally, we state where the
J-type structure has to be truncated for a given time $t$.

  Our new method should therefore lead to more accurate results, and
will allow the construction of much earlier phases of magnetised
shocks.  It shows as well that the criterion used by \citet{L02} to
assess whether steady-states will be of CJ-type (occurrence of a sonic
point in the neutral fluid) has to be revised.  CJ-type steady
states may in fact occur at lower speeds, when velocity recoupling
between neutral and charges enhances magnetic compression near the
piston, and slows down the precursor to the expansion speed of the
J-front.

\subsection{Possible extensions of the method}
  First, let us point out that the time-dependent construction method
derived here relies only on the quasi-steady state assumption for a limited
number of variables, namely velocities, densities, and  magnetic field.
For example, if a set of chemical species can be identified to have
no impact on the dynamics, they can be skipped in the process of building
the truncation radii, and computed only in the last resort. 

Following the same idea, if non-dynamically important species happen
to be non quasi-steady, they can be post-processed in parallel to the
quasi-steady time-evolution with a Lagrangian code. 

  Here we present an algorithm for two kinematic flows (charges and
  neutrals), but the same method may be implemented for more flows.
Especially, the treatment of charged grains could be envisaged, in relation
to the questions raised by \citet{CR02} and \citet{FP03}. The only
caveat is that we do not yet have a consistent check for the validity
of the quasi-steady assumption.

  Finally, our algorithm is straightforward to apply with slowly
changing input conditions $(u,n,B)$ in the piston frame. 
One has only to bear in mind that if these parameters 
change over time-scales much shorter than the crossing time
scale of the shock, then the quasi-steady assumption is very likely to
be violated.

\subsection{Limitations of our method}
 
  Our algorithm is based upon the quasi-steady state
  assumption. However, it will give results with any shock. One
  problem is that we still have no other way to assess the validity of
  the steady-state assumption than computing the time-dependent
  evolution with a fully hydrodynamical code.

  We encountered several cases where this assumption was not realised.
  Strongly unstable shocks like the weakly dissociative ones violate
  strongly this assumption. Fortunately, they seem to happen for a
  very restricted range of parameters. Partly ionising shocks are very
  slightly unstable, and are closer to meet the quasi-steady state
  assumption. They might therefore be accounted for by our
  algorithm. Finally, quite a few magnetised shocks have unstable entrance
  velocities for the J-shock only at early times, and are afterwards
  quasi-steady at all times. These shocks may be as well within reach
  of our algorithm if one is ready to skip the early
  evolution. However, one should always be cautious when a plateau
  with dissociated molecules appears in a steady-state computation.

  An other situation where the quasi-stationary assumption may be
  strongly violated is the case where a dynamically important chemical
  specie is not in a quasi-steady state. This might happen when a
  dominating cooling agent varies on very short time scales.
  Furthermore, diffusion effects, if they turn out to be important,
  will destroy the quasi-steady state as well.

\section{Conclusions}

  In a companion paper (Paper~I), we produced fully time-dependent
  numerical simulations of molecular shocks. 

  In the present paper, we derived new analytical relations valid at
  quasi-steady state, and successfully checked them on our
  simulations. These relations provide useful benchmarks to test
  existing and future multifluid codes.

In light of the simulations run in Paper~I, we investigated carefully
the validity of the quasi-steady state approximation. It was found
that at all times stable shocks could be accounted for by truncated
steady models. We point out as well that caution has to be
  kept regarding the use of steady-state models for dissociative
  velocities.

Finally, we produced a new algorithm based on the quasi-steady state
assumption. With only a steady-state code, this method is
able to compute time-dependent snapshots of shocks in the presence or not
of a magnetic field. Therefore, it brings time-dependence within the reach
of steady models, and should greatly improve the diagnostics of
observed molecular shocks. Furthermore, it provides a way of assessing
the CJ nature of a magnetised shock. Finally, this algorithm can be
extended to many shocks other than molecular, provided that the
quasi-steady state approximation is validated.

\begin{acknowledgements}
We thank the referee (Pr. T.W. Hartquist) for having kindly accepted to 
refer both paper I and paper II at the same time. This work was in part
supported by a European Research \& Training Network (HPRN-CT-20002-00303).
\end{acknowledgements}

\bibliographystyle{aa}

\end{document}